\documentclass[RNAAS]{aastex62}

\begin{document}

\title{VMC J021140.41-735320.4 - an extremely red L dwarf candidate member of the AB Doradus young moving group}

\correspondingauthor{Ralf-Dieter Scholz}
\email{rdscholz@aip.de, lydia.pangsy@arcor.de, mcioni@aip.de}

\author[0000-0002-0894-9187]{Ralf-Dieter Scholz}
\affiliation{Leibniz Institute for Astrophysics Potsdam (AIP),\\
An der Sternwarte 16, 14482 Potsdam, Germany}

\author{Lydia Pangsy}
\affiliation{University of Potsdam (UP), 
Am Neuen Palais 10, 14469 Potsdam, Germany}

\author[0000-0002-6797-696X]{Maria-Rosa L. Cioni}
\affiliation{Leibniz Institute for Astrophysics Potsdam (AIP),\\
An der Sternwarte 16, 14482 Potsdam, Germany}


\section{} 

Free-floating young planetary mass objects and brown dwarfs 
can be directly imaged in deep optical and
infrared sky 
surveys, and their kinematic membership in nearby young moving 
groups (YMGs) provides age determinations
\citep[e.g.][]{2019AJ....157..247R}. 
High proper motions ($\gtrsim100$\,mas/yr) and very red near- and 
mid-infrared colors are typical of nearby young low-mass brown dwarfs
\citep{2017AJ....153..196S}. 
A new tool for discovering such objects is the
Visible and Infrared Survey Telescope for Astronomy 
\citep[VISTA;][]{2006Msngr.126...41E} survey
of the Magellanic Clouds \citep[VMC;][]{2011A&A...527A.116C}
combined with mid-infrared data of the Wide-field Infrared 
Survey Explorer \citep[WISE;][]{2010AJ....140.1868W}. A red (VMC 
photometry $Y=21.284\pm0.125$\,mag, $J=19.856\pm0.050$\,mag, 
$K_s=16.849\pm0.015$\,mag) high
proper motion object, VMC J021140.41-735320.4, was found in  
the Magellanic Bridge region as a result of a master thesis 
\citep[][supervisor M.-R. Cioni]{pangsy19}, involving 16 VMC epochs
(mainly 2011/2012, plus one 2014), one WISE
(2010), and an early epoch (2004) from the South African InfraRed Survey 
Facility \citep[IRSF;][]{2007PASJ...59..615K}.

\begin{figure}[h!] 
\begin{center}
\includegraphics[scale=0.65,angle=0]{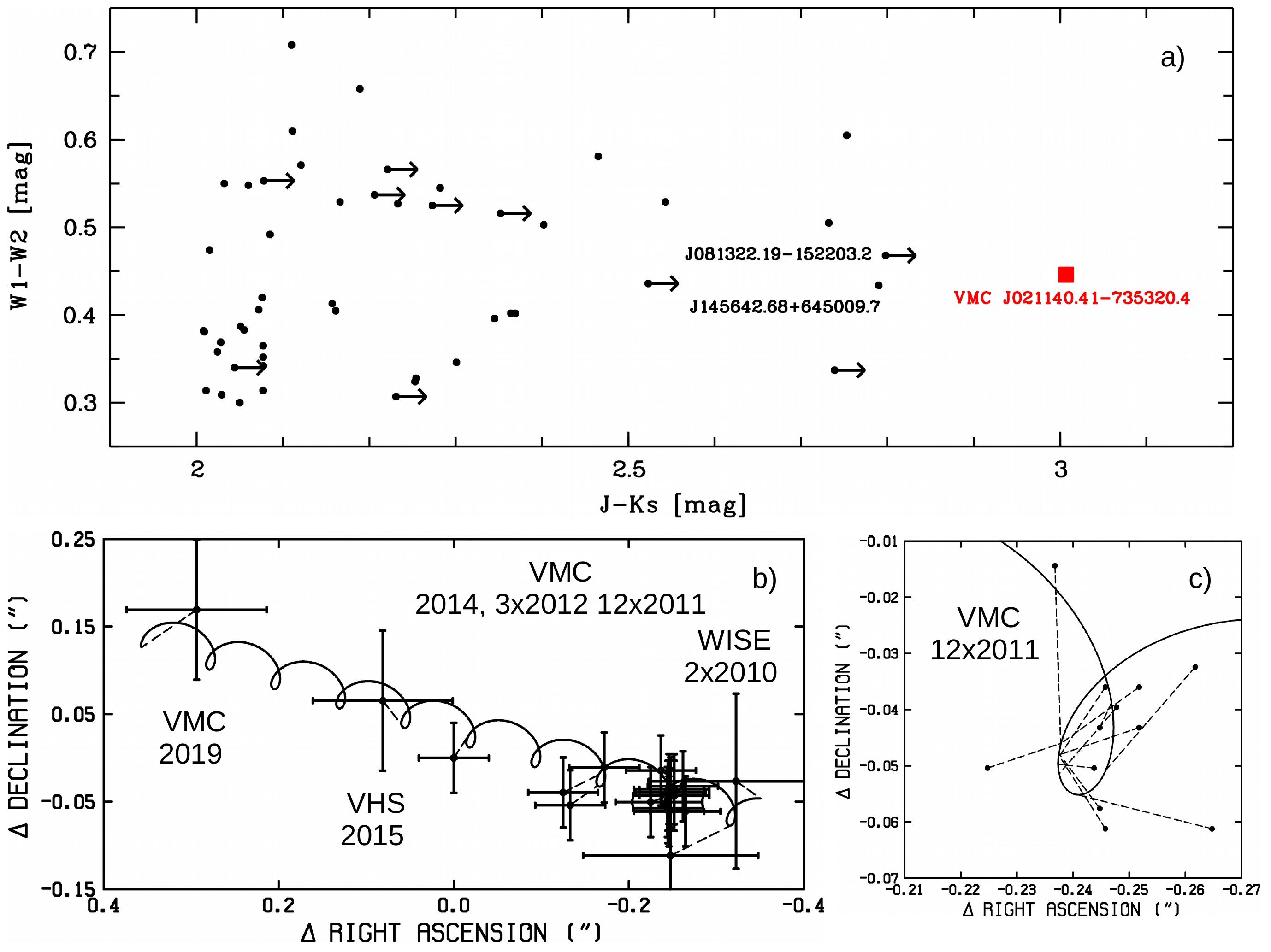}
\caption{a) color-color diagram showing VMC J021140.41-735320.4 and
young L candidates \citep{2017AJ....153..196S}, 
b) astrometric solution\,B, with c) zooming in on 2011.
\label{fig_photpmplx}}
\end{center}
\end{figure}

With its VMC photometry \citep[v1.3;][]{2018MNRAS.474.5459G} and WISE 
all-sky photometry
($W1=15.591\pm0.038$\,mag, $W2=15.145\pm0.062$\,mag) it is even
redder in the near-infrared ($J-K_s=3.0$\,mag) than
the two reddest L dwarfs ($J-K_s \gtrsim 2.8$\,mag, marked in 
Figure~\ref{fig_photpmplx}a) spectroscopically 
classified as ''L7 (sl.red)'' in \citet{2017AJ....153..196S} but
comparable with their mid-infrared color ($W1-W2\approx0.45$\,mag).
Based on their photometric distances ($\approx30$\,pc)
\citep{2017AJ....153..196S} and $JK_sW1W2$ magnitudes, we
estimated for VMC J021140.41-735320.4 a mean photometric distance of 
74\,pc. A similar value of 73\,pc can be
derived using WISEP J004701.06+680352.1 (L7.5p red), one 
of the few known peculiar (red) late-L dwarfs \citep{2017MNRAS.469..401S}
with a trigonometric parallax \citep{2016ApJ...833...96L}. However,
the latter and the recently discovered nearby young low-gravity L7 dwarf 
2MASS J07555430-3259589 \citep{2018RNAAS...2...33S,2019RNAAS...3...30L}
have much smaller near-infrared ($J-K_s=2.5-2.7$\,mag) 
but larger mid-infrared color indices ($W1-W2\approx0.60$\,mag). 
Because we lack comparison objects whose colors exactly match those of
VMC J021140.41-735320.4, its photometric distance remains
uncertain.

In a preliminary proper motion and parallax determination using 
software of \citet{2001AAS...198.4709G}, we included additional positions
from a 2019 VMC observation (under bad sky conditions), 
VISTA hemisphere survey \citep[VHS;][epoch 2015]{2013Msngr.154...35M},
WISE post-cryo (2010), and NEOWISE data (10 epochs 2014-2018).
Using all 31 positions, with different weights according to
assumed errors of 40\,mas (VMC), 80\,mas (VMC-2019 and VHS), 100\,mas (WISE),
200\,mas (NEOWISE), and 250\,mas (IRSF), we obtained solution\,A:
$\mu_{\alpha}\cos{\delta}=+73.3\pm3.2$\,mas/yr,
$\mu_{\delta}=+22.0\pm3.2$\,mas/yr, and
$\pi=20.4\pm7.4$\,mas.
Excluding the less precise IRSF and NEOWISE data, 
our prefered solution\,B from 20 positions 
(Figure~\ref{fig_photpmplx}b,c) was:
$\mu_{\alpha}\cos{\delta}=+74.3\pm3.4$\,mas/yr,
$\mu_{\delta}=+22.3\pm3.4$\,mas/yr, and
$\pi=21.0\pm7.1$\,mas,
consistent with solution\,A but yielding a slightly more significant 
parallax.
Using BANYAN (http://www.exoplanetes.umontreal.ca/banyan/)
of \citet{2018ApJ...856...23G}, we found 
66.6\% (solution\,A) and 67.8\% (solution\,B) membership probabilities
of VMC J021140.41-735320.4 in the AB Doradus YMG \citep{2004ApJ...613L..65Z}
that may be coeval ($\approx$125\,Myr) with the Pleiades
\citep{2013ApJ...766....6B}.
With its trigonometric distance of $\approx$48\,pc, VMC J021140.41-735320.4 
has an absolute magnitude of $M_{K_s}\approx13.4$\,mag, 
in agreement with
the planetary mass companion ($M_{K_s}=13.35\pm0.18$\,mag) of the AB Doradus 
YMG member candidate 2MASS J22362452+4751425 \citep{2017AJ....153...18B}.
Spectroscopy and more accurate astrometry are needed to confirm
VMC J021140.41-735320.4 as an AB Doradus YMG member.

\acknowledgments

MRC acknowledges support from the European Research Council
(grant No 682115).

\end{document}